\documentclass{article}

\newtheorem{theorem}{Theorem}

\title{Teleparallelism, modified Born-Infeld nonlinearity\\
and space-time as a micromorphic ether.}
\author{ Jan J. S\l awianowski \\
Institute of Fundamental Technological Research, \\
Polish Academy of Sciences, \\
21 \'{S}wi\c{e}tokrzyska str., 00-049 Warsaw, Poland \\
e-mail: jslawian@ippt.gov.pl}

\begin{document}

\maketitle

\begin{abstract}
Discussed are field-theoretic models with degrees of freedom described by the
$n$-leg field in an $n$-dimensional "space-time" manifold. Lagrangians are
generally-covariant and invariant under the internal group GL$(n,{\bf R})$. It is
shown that the resulting field equations have some correspondence with Einstein
theory and possess homogeneous vacuum solutions given by semisimple Lie group spaces
or their appropriate deformations. There exists a characteristic link with the
generalized Born-Infeld type nonlinearity and relativistic mechanics of structured
continua. In our model signature is not introduced by hands, but is given by
integration constants for certain differential equations.
\end{abstract}


\section{Introduction}

The model suggested here has several roots and arose from some very peculiar and
unexpected convolution of certain ideas and physical concepts seemingly quite remote
from each other. In a sense it unifies generalized Born-Infeld type nonlinearity,
tetrad approaches to gravitation, Hamiltonian systems with symmetries (mainly with
affine symmetry; so-called affinely-rigid body), generally-relativistic spinors and
motion of generalized relativistic continua with internal degrees of freedom
(relativistic micromorphic medium, a kind of self-gravitating microstructured "ether"
generalizing the classical Cosserat continuum). The first two of mentioned topics
(Born-Infeld, tetrad methods) were strongly contributed by Professor Jerzy
Pleba\'{n}ski, cf. e.g. \cite{16,18}. The same concerns spinor theory \cite{17}. My
"micromorphic ether", although in a rather very indirect way, is somehow related to
the problem of motion in general relativity; the discipline also influenced in a
known way by J.~Pleba\'{n}ski \cite{13}. There exist some links between generalized
Born-Infeld nonlinearity and the modern theory of strings, membranes and p-branes.
Geometrically this has to do with the theory of minimal surfaces \cite{9}.

\section{Born-Infeld motive}

Let us begin with the Born-Infeld motive of our study. No doubt,
linear theories with their superposition principle are in a sense
the simplest models of physical phenomena. Nevertheless, they are
too poor to describe physical reality in an adequate way. They are
free of the essential self-interaction. In linear electrodynamics
stationary centrally symmetric solutions of field equations are
singular at the symmetry centre and their total field energy is
infinite. If interpreting such centres as point charges one
obtains infinite electromagnetic masses, e.g. for the electron. In
realistic field theories underlying elementary particle physics
one usually deals with polynomial nonlinearity, e.g. the quartic
structure of Lagrangians is rather typical. Solitary waves
appearing in various branches of fundamental and applied physics
owe their existence to various kinds of nonlinearity, very often
nonalgebraic ones. General relativity is nonlinear (although
quasilinear at least in the gravitational sector) and its
equations are given by rational functions of field variables,
although Lagrangians themselves are not rational. In Einstein
theory one is faced for the first time with a very essential
non-linearity which is not only nonperturbative (it is not a small
nonlinear correction to some dominant linear background) but is
also implied by the preassumed symmetry conditions, namely by the
demand of general covariance. Indeed, any Lagrangian theory
invariant under the group of all diffeomorphisms must be nonlinear
(although, like Einstein theory may be quasilinear). Nonlinearity
of non-Abelian gauge theories is also due to the assumed symmetry
group. In our mechanical study of affinely-rigid bodies \cite{20}
nonlinearity of geodetic motion was also due to the assumption of
invariance under the total affine group. This, by the way,
established some link with the theory of integrable lattices.

The original Born-Infeld nonlinearity had a rather different
background and was motivated by the mentioned problems in Maxwell
electrodynamics. There was also a tempting idea to repeat the
success of general relativity and derive equations of charges
motion from the field equations. Unlike the problem of infinities,
which was in principle solved, the success in this respect was
rather limited. The reason is that in general relativity the link
between field equations and equations of motion is due not only to
the nonlinearity itself (which is, by the way, necessary), but
first of all to Bianchi identities. The latter follow from the
very special kind of nonlinearity implied by the general
covariance.

As we shall see later on, some kinds of generalized Born-Infeld nonlinearities also
may be related to certain symmetry demands. But for us it is more convenient to begin
with some apparently more formal, geometric aspect of "Born-Infeld-ism".

In linear theories Lagrangians are built in a quadratic way from the field variables
$\Psi$. Thus, they may involve $\Psi \Psi$-terms algebraically quadratic in $\Psi,\
\partial \Psi \partial \Psi$-terms algebraically quadratic in derivatives of $\Psi$,
and $\Psi \partial \Psi$-terms bilinear in $\Psi$ and $\partial
\Psi$; everything with constant coefficients. In any case, the
dependence on derivatives, crucial for the structure of field
equations is polynomial of at most second order in $\partial \Psi$
(linear with $\Psi$-coefficients in the case of fermion fields).

But there is also another, in a sense opposite pole of
mathematical simplicity of Lagrangians. By its very geometrical
nature, Lagrangian $L$ is a scalar Weyl density of weight one; in
an $n$-dimensional orientable manifold it may be represented by a
differential $n$-form locally given by: $\pounds= L(\Psi,\partial
\Psi)dx^{1}\wedge \ldots \wedge dx^{n}.$ But as we know, there is
a canonical way of constructing such densities: just taking square
roots of the moduli of determinants of second-order covariant
tensors,
\begin{equation}\label{1}
L=\sqrt{|{\rm det}[L_{ij}]|},
\end{equation}
or rather constant multiples of this expression, when some
over-all negative sign may occur. In the sequel the square-rooted
tensor will be referred to as the Lagrange tensor, or tensorial
Lagrangian. In general relativity and in all field theories
involving metric tensor $g$ on the space-time manifold $M$, $L$ is
factorized in the following way: $L=\Lambda (\Psi,\partial
\Psi)\sqrt{|{\rm det}[g_{ij}]|},$ where $\Lambda$ is a scalar
expression. Here the square-rooted metric $g_{ij}$ offers the
canonical scalar density. In linear theories for fields $\Psi$
considered on a fixed metrical background $g,\ \Lambda$ is
quadratic (in the aforementioned sense) in $\Psi$. In quasilinear
Einstein theory, where in the gravitational sector $\Psi$ is just
$g$ itself, one uses the Hilbert Lagrangian proportional to
$R[g]\sqrt{|g|}$ ($R[g]$ denoting obviously the scalar curvature
of $g$ and $|g|$ being an obvious abbreviation for the modulus of
${\rm det}[g_{ij}]$). Obviously, Lagrangians of linear and
quasilinear theories also may be written in the form (\ref{1}),
however, this representation is extremely artificial and
inconvenient. For example, for the Hilbert Lagrangian we have $
L={\rm sign}R\sqrt{|{\rm det}[R^{2/n}g_{ij}]|}, $ i.e. locally we
can write $ L_{ij}=|R|^{2/n}g_{ij};\ L_{ij}=\sqrt{|R|}g_{ij}\ {\rm
if}\ n=4.$

One can wonder whether there exist phenomena reasonably and in a
convenient way described just in terms of (\ref{1}). It is natural
to expect that the simplest models of this type will correspond to
the at most quadratic dependence of the tensor $L_{ij}$ on field
derivatives. Unlike the linear and quasilinear models, now the
theory structure will be lucid just on the level of $L_{ij}$.
Among all possible nonlinear models such ones will be at the same
time quite nonperturbative but also in a sense similar to the
linear and quasilinear ones.

The historical Born-Infeld model \cite{2} is exceptional in that the Lagrange tensor
is linear in field derivatives; Lagrangian is
\begin{equation}\label{2}
L=-\sqrt{|{\rm det}[bg_{ij}+F_{ij}]|}+b^{2}\sqrt{|{\rm det}[g_{ij}]|},
\end{equation}
where $F_{ij}=A_{j,i}-A_{i,j}$ is the electromagnetic field strength, $A_{i}$ is the
covector potential and the constant $b$ is responsible for the saturation phenomenon;
it determines the maximal attainable field strength. The field dynamics is encoded in
the first term. The second one, independent of $F$, fixes the energy scale:
Lagrangian and energy are to vanish when $F$ vanishes. Therefore, up the minus sign
preceding the square root, we have $L_{ij}=bg_{ij}+F_{ij}$. For weak fields, e.g. far
away from sources, $L$ asymptotically corresponds with the quadratic Maxwell
Lagrangian. And all singularities of Maxwell theory are removed --- static
spherically symmetric solutions are finite at the symmetry centre (point charge) and
the electromagnetic mass is finite. The finiteness of solutions is due to the
saturation effect. $L$ has a differential singularity of the type $\sqrt{0}$ when the
field is so strong that the determinant of $\left[bg_{ij}+F_{ij}\right]$ vanishes.
Such a situation is singular-repulsive just as $v=c$ situation for the relativistic
particle, where the interaction-free Lagrangian is given by
$L=-mc^{2}\sqrt{1-v^{2}/c^{2}}$ (in three-dimensional notation). The classical
Born-Infeld theory is in a sense unique, exceptional among all a priori possible
models of nonlinear electrodynamics \cite{1,18}. It is gauge invariant, the energy
current is not space-like, energy is positively definite, point charges have finite
electromagnetic masses and there is no birefringence. There exist solutions of the
form of plane waves combined with the constant electromagnetic field; in particular,
solitary solutions may be found \cite{1}.

Although the amazing success of quantum field theory and
renormalization techniques (even classical ones, as developed by
Dirac) for some time reduced the interest in Born-Infeld theory,
nowadays this interest is again growing on the basis of new
motivation connected, e.g. with strings, p-branes, alternative
approaches to gravitation, etc. \cite{5,6,7}.

Linearity of $L_{ij}$ in field derivatives is an exceptional feature of
electrodynamics among all models developed in the Born-Infeld spirit. Usually
$L_{ij}$ must be quadratic in derivatives because of purely geometric reasons. For
example, let us consider the scalar theory of light, neglecting the polarization
phenomena. In linear theory one uses then the real scalar field $\Psi$ ruled by the
d'Alembert Lagrangian $ L=g^{ij}\Psi_{,i}\Psi_{,j}\sqrt{|g|}.$ The only natural
"Born-Infeld-ization" of this scheme is based on
\begin{equation}\label{3}
L=-\sqrt{|{\rm det}[bg_{ij}+\Psi_{,i}\Psi_{,j}]|}+b^{2}\sqrt{|{\rm det}[g_{ij}]|},
\end{equation}
thus, as a matter of fact $L_{ij}$ is quadratic in field derivatives $\partial \Psi$.
It is interesting that such a model gives for the stationary spherically symmetric
solutions in Minkowski space the formula which is exactly identical with that for the
scalar potential $\varphi=A_{0}$ in the usual Born-Infeld model, namely, the
expression $ f(r)=\sqrt{Ab}\int_{0}^{r}du/\sqrt{A+u^{4}},$ where $A$ denotes the
integration constant (related to the value of point charge producing the field). Let
us mention, incidentally, that such a scalar Born-Infeld model was successfully
applied in certain problems of nonlinear optics. Therefore, the very use of $L_{ij}$
quadratic in derivatives does not seem to violate philosophy underlying the
Born-Infeld model. There are also other arguments. Born-Infeld theory explains point
charges in a very nice way as "regular singularities", but is not well-suited to
describing interactions with autonomous external sources, e.g. with the charged
(complex) Klein-Gordon or Dirac fields. Combining in the usual way Born-Infeld
Lagrangian with expressions describing matter fields and the mutual interactions one
obtains equations involving complicated nonrational expressions. It seems much more
natural to use the expression (\ref{1}) with $L_{ij}$, e.g. of the form: $
L_{ij}=\alpha g_{ij}+\kappa \overline{\Psi}\Psi g_{ij}+
bF_{ij}+cD_{i}\overline{\Psi}D_{j}\Psi, $ where $\alpha,\kappa,b,c$ are constants and
$D_{j}\Psi=\Psi_{,j}+ieA_{j}\Psi$ (electromagnetic covariant derivatives). Here
$\Psi$ denotes the complex scalar field and for weak fields we obtain the usual
mutually coupled Maxwell-Klein-Gordon system. The same may be done obviously for
field multiplets and for fermion fields. Such a model has a nice homogeneous
structure and the field equations are rational in spite of the square-root expression
used in Lagrangian.

Once accepting $L_{ij}$ quadratic in derivatives we can also think
about admitting such models also for the pure electromagnetic
field, e.g.
\begin{equation}\label{4}
L_{ij}=\alpha g_{ij}+\beta F_{ij}+ \gamma g^{kl}F_{ik}F_{lj}+\delta
g^{kr}g^{ls}F_{kl}F_{rs}g_{ij},
\end{equation}
where $\alpha,\beta,\gamma,\delta$ are constants. The terms
quadratic in $F$ in (\ref{4}) are known as contributions to the
energy-momentum tensor of the Maxwell field.

If we try to construct Born-Infeld-like models for non-Abelian
gauge fields, then the quadratic structure of $L_{ij}$ is as
unavoidable as in scalar electrodynamics, e.g.
\begin{equation}\label{5}
L_{ij}=\alpha g_{ij}+\gamma g^{kl}F^{K}{}_{ik}F^{L}{}_{lj}h_{KL}
\end{equation}
is the most natural expression. Obviously, $\alpha,\gamma$ are
constants, $F^{K}{}_{ij}$ are strength of the gauge fields and
$h_{KL}$ is the Killing metric on the gauge group Lie algebra.

Lagrangians (\ref{1}) may be modified by introducing "potentials",
i.e. scalar $\Psi$-dependent multipliers at the square-root
expression, or at $L_{ij}$ itself, or finally, at the determinant.
However, the less number of complicated and weakly-motivated
corrections of this type, the more aesthetic and convincing is the
dynamical hypothesis contained in $L$.

It is worthy of mentioning that the scalar Born-Infeld models with
quadratic $L_{ij}$ have to do with the theory of minimal surfaces
\cite{9} and with some interplay between general covariance and
internal symmetry. Namely, we can consider scalar fields $\Psi$ on
$M$ with values in some linear space $W$ of dimension $m$ higher
than $n={\rm dim}M$. Let $W$ be endowed with some
(pseudo)Euclidean metric $h\in W^{*}\otimes W^{*}$. We could as
well consider pseudo-Riemannian structure as a target space,
however, now we prefer to concentrate on the simplest model. If
$M$ is structureless, then the only natural possibility of
constructing Lagrangian invariant under rotations O$(W,\eta)$ and
under Diff$M$ (generally-covariant) is to take the pull-back
Lagrange tensor
$L_{ij}=g_{ij}=h_{KL}\Psi^{K}{}_{,i}\Psi^{L}{}_{,j}.$ If $h$ is
Euclidean, this means that we search minimal surfaces in $W$; $M$
is used as a merely parametrization. Field equations have the form
$g^{ij}\nabla_{i}\nabla_{j}\Psi^{K}=0,\ K=\overline{1,m},$ where
the covariant differentiation is meant in the Levi-Civita
$g$-sense. We can fix the coordinate gauge by putting
$\left((1/2)g^{ij}g^{ab}-g^{ia}g^{jb}\right)g_{ab,i}=0,$ e.g.
making the assumption $\Psi^{i}=x^{i},\ i=\overline{1,n}$, i.e.
identifying $n$ of the fields $\Psi^{K}$ with $M$-coordinates
themselves. Then the gauge-free content of our field equations is
given by: $g^{ij}\Psi^{\Sigma}{}_{,ij}=0,\
\Sigma=\overline{n+1,m}. $ These equations follow from the
effective Lagrange tensor
\begin{equation}\label{6}
L^{\rm eff}{}_{ij}=h_{ij}+2h_{\Sigma (i}\Psi^{\Sigma}{}_{,j)}+h_{\Sigma \Lambda}
\Psi^{\Sigma}{}_{,i}\Psi^{\Lambda}{}_{,j}.
\end{equation}
Here we easily recognize something similar to (\ref{4}), i.e.
second-order polynomial in derivatives with the effective
background metric $h_{ij}$ in $M$. If $h$ has the block structure
with respect to $(\Sigma,i)$-variables, then there are no
first-order terms, just as in (\ref{3}),(\ref{5}). It is seen that
the "almost classical" Born-Infeld form with the effective metric
on $M$ may be interpreted as a gauge-free reduction of
generally-covariant dynamics in $M$ with some internal symmetries
in the target space $W$. One can also multiply the corresponding
Lagrangians by some "potentials" depending on the $h$-scalars
built of $\Psi$, however with the provisos mentioned above.
Surprisingly enough, such scalar models describe plenty of
completely different things, e.g. soap and rubber films, geodetic
curves, relativistic mechanics of point particles, strings and
p-branes, minimal surfaces and Jacobi-Maupertuis variational
principles. There were also alternative approaches to gravitation
based on such models \cite{14}.

\section{Tetrads, teleparallelism and internal affine\\ symmetry}

There are various reasons for using tetrads in gravitation theory, in particular for
using them as gravitational potentials, in a sense more primary then the metric
tensor \cite{15}. First of all they provide local reference frames reducing the metric
tensor to its Minkowskian shape. They are unavoidable when dealing with spinor fields
in general relativity. This has to do with the curious fact that $\overline{{\rm
GL}^{+}(n,{\bf R})}$, the universal covering group of ${\rm GL}^{+}(n,{\bf R})$, is
not a linear group (has no faithful realization in terms of finite matrices). Also
the gauge approaches to gravitation (SL$(2,{\bf C})$-gauge, Poincar\'{e} gauge
models) are based on the use of tetrad fields. And even in standard Einstein theory
the tetrad formulation enables one to construct first-order Lagrangians which are
well-defined scalar densities of weight one. If one uses the metric field as a
gravitational potential, the Hilbert Lagrangian is, modulo the cosmological term, the
only possibility within the class of essentially first-order variational principles.
Unlike this, the tetrad degrees of freedom admit a wide class of nonequivalent
variational principles. Some of them were expected to overwhelm singularities
appearing in Einstein theory.

Let us begin with introducing necessary mathematical concepts. It
is convenient to consider a general "space-time" manifold $M$ of
dimension $n$ and specify to $n=4$ only on some finite stage of
discussion. The principal fibre bundle of linear frames will be
denoted by $\pi : FM \rightarrow M$ and its dual bundle of
co-frames by $\pi^{*}: F^{*}M \rightarrow M$. The duality between
frames and co-frames establishes the canonical diffeomorphism
between $FM$ and $F^{*}M$. The co-frame dual to $e=(\ldots
,e_{A},\ldots)$ will be denoted by $\widetilde{e}=(\ldots,
e^{A},\ldots)$; by definition $\langle
e^{A},e_{B}\rangle=\delta^{A}{}_{B}$. When working in local
coordinates $x^{i}$ we use the obvious symbols $e^{i}{}_{A},\
e^{A}{}_{i}$, omitting the tilde-sign at the co-frame. Therefore,
$e^{A}{}_{i}e^{i}{}_{B}=\delta^{A}{}_{B},\
e^{i}{}_{A}e^{A}{}_{j}=\delta^{i}{}_{j}$. The structure group
GL$(n,{\bf R})$ acts on $FM,\ F^{*}M$ in a standard way, i.e. for
any $L\in {\rm GL}(n,{\bf R}):$ $e \mapsto
eL=(\ldots,e_{A},\ldots)L=(\ldots,e_{B}L^{B}{}_{A},\ldots),\
\widetilde{e} \mapsto \widetilde{e}L=(\ldots,e^{A},\ldots)L=
(\ldots,L^{-1}{}^{A}{}_{B}e^{B},\ldots).$ Fields of (co-)frames
((co-)tetrads when $n=4$) are sufficiently smooth cross-sections
of $F^{*}M$, respectively $FM$ over $M$. They are affected by
elements of GL$(n,{\bf R})$ pointwise, according to the above
rule. In gauge models of gravitation one must admit local, i.e.
$x$-dependent action of GL$(n,{\bf R})$. Any field $M \ni x\mapsto
L(x) \in {\rm GL}(n,{\bf R})$ acts on cross-section $M \ni x
\mapsto e_{x} \in FM$ according to the rule: $
\left(eL\right)_{x}=e_{x}L(x).$ Obviously, for any $x \in M,\
e_{x} \in \pi^{-1}(x)$ may be identified with a linear isomorphism
of ${\bf R}^{n}$ onto the tangent space $T_{x}M$; similarly,
$\widetilde{e}_{x} \in \pi^{*}{}^{-1}(x)$ is an ${\bf
R}^{n}$-valued form on $T_{x}M$. Therefore, the field of co-frames
is an ${\bf R}^{n}$-valued differential one-form on $M$. In
certain problems it is convenient to replace ${\bf R}^{n}$ by an
abstract $n$-dimensional linear space $V$. The reason is that
${\bf R}^{n}$ carries plenty of structures sometimes considered as
canonical (e.g. the Kronecker metric) and this may lead to false
ideas.

In the sequel we shall need some byproducts of the field of frames. If $\eta$ is a
pseudo-Euclidean metric on ${\bf R}^{n}$ (on $V$), then the Dirac-Einstein metric
tensor on $M$ is defined as: $ h[e,\eta]=\eta_{AB}e^{A}\otimes e^{B},\
h_{ij}=\eta_{AB}e^{A}{}_{i}e^{B}{}_{j}. $ In general relativity $n=4$ and
$[\eta_{AB}]={\rm diag}(1,-1,-1,-1)$. Obviously, the prescription for $e\mapsto
h[e,\eta]$ is invariant under the local action of the pseudo-Euclidean group
O$(n,\eta)$. In general relativity it is the Lorentz group O$(1,3)$ that is used as
internal symmetry. The metric $\eta$, or rather its signature, is an absolute element
of the theory.

The field of frames gives rise to the teleparallelism connection
$\Gamma_{\rm tel}[e]$; it is uniquely defined by the condition
$\nabla e_{A}=0,\ A=\overline{1,n}.$ In terms of local
coordinates: $\Gamma^{i}{}_{jk}=e^{i}{}_{A}e^{A}{}_{j,k}.$
Obviously, its curvature tensor vanishes and the parallel
transport of tensors consists in taking in a new point the tensor
with the same anholonomic $e$-components. The prescription $e
\mapsto \Gamma (e)$ is globally GL$(n,{\bf R})$-invariant,
$\Gamma[eA]=\Gamma[e],\ A \in {\rm GL}(n,{\bf R}).$ The torsion
tensor of $\Gamma_{\rm tel}$, $ S[e]^{i}{}_{jk}=\Gamma_{\rm
tel}{}^{i}{}_{[jk]}=
(1/2)e^{i}{}_{A}\left(e^{A}{}_{j,k}-e^{A}{}_{k,j}\right)$ may be
interpreted as an invariant tensorial derivative of the field of
frames. It is directly related to the non-holonomy object $\gamma$
of $e$, $
S^{i}{}_{jk}=\gamma^{A}{}_{BC}e^{i}{}_{A}e^{B}{}_{j}e^{C}{}_{k},\
\left[e_{A},e_{B}\right]=\gamma^{C}{}_{AB}e_{C}$ (as usual,
$[u,v]$ denotes the Lie bracket of vector fields $u,v$).

In general relativity tetrad field is interpreted as a gravitational potential; the
space-time metric $h[e,\eta]$ is a secondary quantity. When expressed through $e$,
Hilbert Lagrangian may be invariantly reduced to some well-defined scalar density of
weight one and explicitly free of second derivatives. Indeed, one can show that
\begin{equation}\label{7}
L_{\rm H}=R
[h[e]]\sqrt{|h|}=\left(J_{1}+2J_{2}-4J_{3}\right)\sqrt{|h|}+
4(S^{a}{}_{ab}h^{bi}\sqrt{|h|})_{,i},
\end{equation}
where $|h|$ is an abbreviation for $|{\rm det}\left[h[e]_{ij}\right]|$ and
$J_{1}=h_{ai}h^{bj}h^{ck}S^{a}{}_{bc}S^{i}{}_{jk},\
J_{2}=h^{ij}S^{a}{}_{ib}S^{b}{}_{ja},\ J_{3}=h^{ij}S^{a}{}_{ai}S^{b}{}_{bj}$ are
Weitzenb\"ock invariants built quadratically of $S$. They are invariant under the
global action of O$(1,3)$ on $e$. The last term in (\ref{7}) is a well-defined scalar
density and the divergence of some vector density of weight one. It absorbs the
second derivatives of $e$. Therefore, Hilbert Lagrangian is equivalent to the first
term in (\ref{7}),
\begin{equation}\label{8}
L_{\rm{H-tel}}:=
L_{1}+2L_{2}-4L_{3}=\left(J_{1}+2J_{2}-4J_{3}\right)\sqrt{|h|}.
\end{equation}
It is invariant under the local action of O$(1,3)$ modulo appropriate divergence
corrections. And resulting field equations for $e$ are exactly Einstein equations
with $h[e,\eta]$ substituted for the metric tensor. In this sense one is dealing with
different formulation of the same theory. Obviously, due to the mentioned local
O$(1,3)$-invariance, the tetrad formulation involves more gauge variables.

The use of tetrads as fundamental fields opens the possibility of formulating more
general dynamical models. The simplest modification consists in admitting general
coefficients at three terms of (\ref{8}),
\begin{equation}\label{9}
L=c_{1}L_{1}+c_{2}L_{2}+c_{3}L_{3}.
\end{equation}
When the ratio $c_{1}:c_{2}:c_{3}$ is different than $1:2:(-4)$, the resulting model
loses the local O$(1,3)$-invariance and is invariant only under the global action of
O$(1,3)$. The whole tetrad $e$ becomes a dynamical variable, whereas in (\ref{8})
everything that does not contribute to $h[e,\eta]$ is a pure gauge. Models based on
(\ref{9}) were in fact studied and it turned out that in a certain range of
coefficients $c_{1},\ c_{2},\ c_{3}$ their predictions agree with those of Einstein
theory and with experiment. One can consider even more general models with
Lagrangians non-quadratic in $S$:
\begin{equation}\label{10}
L(S,h)=g(S,h)\sqrt{|h|},
\end{equation}
where $g$ is arbitrary scalar intrinsically built of $S,h,$ e.g.
some nonlinear function of Weitzenb\"{o}ck invariants. The
resulting theories are not quasilinear any longer. There were some
hopes to avoid certain non-desirable infinities by appropriate
choice of $g$ (people were then afraid of singularities, nowadays
they love them). To write down field equations in a concise form
it is convenient to introduce two auxiliary quantities
$H_{i}{}^{jk}:=\partial L/\partial S^{i}{}_{jk}
=e^{A}{}_{i}H_{A}{}^{jk}= e^{A}{}_{i}\partial L/\partial
e^{A}{}_{j,k},\ Q^{ij}:=\partial L/\partial h_{ij}$ referred to,
respectively, as a field momentum and Dirac-Einstein stress. They
are tensor densities of weight one. One can show that equations of
motion have the form:
$K_{i}{}^{j}:=\nabla_{k}H_{i}{}^{jk}+2S^{l}{}_{lk}H_{i}{}^{jk}-2h_{ik}Q^{kj}=0.
$ The covariant differentiation is meant here in the
$e$-teleparallelism sense.

To the best of our knowledge, all teleparallelism models of
gravitation belonged to the above described class. They are
invariant under global action of O$(n,\eta)$ (i.e. O$(1,3)$ in the
physical four-dimensional case). Let us observe, however, that
there are some fundamental philosophical objections concerning
this symmetry. The corresponding local symmetry in Einstein theory
was well-motivated. It simply reflected the fact that the tetrad
field was a merely nonholonomic reference frame, something without
a direct dynamical meaning. It was only its metrical aspect
$h[e,\eta]$ that was physically interpretable. If we once decide
seriously to make the total $e$ a dynamical quantity, the global
O$(e,\eta)$-symmetry evokes some doubts. Why not the total
GL$(n,{\bf R})$-symmetry? Why to introduce by hands the
Minkowskian metric $\eta$ to ${\bf R}^{n}$, the internal space of
tetrad field? Such questions become very natural when, as
mentioned above, we use an abstract linear space $V$ instead of
${\bf R}^{n}$. From the purely kinematical point of view the most
natural group is GL$(V)$. It seems rather elegant to use a bare,
amorphous linear space $V$ than to endow it a priori with
geometrically nonmotivated absolute element $\eta \in V^{*}
\otimes V^{*}$. To summarize: when one gives up the local Lorentz
symmetry O$(V,\eta)$, then it seems more natural to use the global
GL$(V)$ than global O$(V,\eta)$. Then $L$ in (\ref{10}) does not
depend on $h$ and our field equations for Lagrangians $L(S)$ have
the following general form:
\begin{equation}\label{11}
K_{i}{}^{j}:=\nabla_{k}H_{i}{}^{jk}+2S^{l}{}_{lk}H_{i}{}^{jk}=0.
\end{equation}
If the model is to be generally-covariant, that we always assume,
then some Bianchi-type identities imply that $L$ is an $n$-th
order homogeneous function of $S$, $S^{i}{}_{jk}\partial
L/\partial S^{i}{}_{jk}=S^{i}{}_{jk}H_{i}{}^{jk}=nL,$ i.e.
$L(\lambda S)=\lambda^{n}L(S)$ for any $\lambda >0$. This is a
kind of generalized Finsler structure.

If we search for models with internal linear-conformal symmetry
${\bf R}^{+}{\rm O}(V,\eta)$$=e^{\bf R}{\rm O}(V,\eta)$, then $L$
must be homogeneous of degrees $0$ in $h$, $h_{ij}\partial
L/\partial h_{ij}=h_{ij}Q^{ij}=0,$ i.e. $L(S,\lambda h)=L(S,h)$
for any $\lambda \in {\bf R}^{+}$.

The simplest GL$(n,{\bf R})$-invariant (GL$(V)$-invariant) and generally-covariant
models have the following generalized Born-Infeld structure: $L=\sqrt{|{\rm
det}[L_{ij}]|}$ with the Lagrange tensor quadratic in derivatives:
\begin{equation}\label{12}
L_{ij}=4\lambda S^{k}{}_{im}S^{m}{}_{jk}+4\mu S^{k}{}_{ik}S^{m}{}_{jm}+ 4\nu
S^{k}{}_{lk}S^{l}{}_{ij},
\end{equation}
where $\lambda,\mu,\nu$ are real constants. One can in principle
complicate them and make more general multiplying the above
Lagrangian $L$, or Lagrange tensor $L_{ij}$, or the under-root
expression (to some extent the same procedure) by a function of
some basic GL$(V)$-invariant and generally-covariant scalars built
of $S$. All such scalars are zeroth-order homogeneous functions of
$S$.

The first two terms of (\ref{12}) are symmetric and may be considered as a candidate
for the metric tensor of $M$ built of $e$ in a globally GL$(V)$-invariant and
generally-covariant way:
\begin{equation}
g_{ij}=\lambda \gamma_{ij}+\mu \gamma_{i}\gamma_{j}=4\lambda S^{k}{}_{im}S^{m}{}_{jk}
+4\mu S^{k}{}_{ik}S^{m}{}_{jm}.
\end{equation}
The best candidate is the dominant term $\gamma_{ij}$ built of $S$ according to the
Killing prescription.

The mentioned scalar potentials used for multiplying (\ref{12})
may be built of expressions like
$\gamma_{il}\gamma^{jm}\gamma^{kn}S^{i}{}_{jk}S^{l}{}_{mn},\
\gamma^{ij}S^{k}{}_{ik}S^{m}{}_{jm},\
\Gamma^{i}{}_{j}\Gamma^{j}{}_{k}\ldots\Gamma^{l}{}_{m}\Gamma^{m}{}_{i},$
etc., where $\Gamma_{ij}:=4S^{k}{}_{lk}S^{l}{}_{ij},\
\Gamma^{i}{}_{j}=\gamma^{im}\Gamma_{mj}$ and $\gamma^{ij}$ is
reciprocal to $\gamma_{ij}$,
$\gamma^{ik}\gamma_{kj}=\delta^{i}{}_{j};$ we assume it does
exist.

No doubts, the simplest and maximally "Born-Infeld-like" are models without such
scalar potential terms, with the Lagrange tensor (\ref{12}) quadratic in derivatives.

Due to the very strong nonlinearity, it would be very difficult to perform in all
details the Dirac analysis of constraints resulting from the Lagrangian singularity.
Nevertheless, the primary and secondary constraints may be explicitly found. If the
problem is formulated in $n$ dimensions and all indices both holonomic and
nonholonomic are written in the convention $K,i=\overline{0,n-1}$ (zeroth variable
referring to "time"), then primary constraints, just as in electrodynamics, are given
by $\pi^{0}{}_{K}=0,$ where $\pi^{i}{}_{K}$ are densities of canonical momenta
conjugated to "potentials" $e^{K}{}_{i}$. Thus, there are a priori $n$ redundant
variables among $n^{2}$ quantities $e^{i}{}_{K}$ and they may be fixed by coordinate
conditions, like, e.g. $e^{i}{}_{K}=\delta^{i}{}_{K}$ for some fixed value of $K$ or
$e^{K}{}_{i}{}^{,i}=e^{K}{}_{i,j}g^{ji}=0$ (Lorentz transversality condition).
Secondary constraints are related to the field equations free of second "time"
derivatives, these may be shown to be:
$K_{i}{}^{0}=\nabla_{j}H_{i}{}^{0j}+2S^{k}{}_{kj}H_{i}{}^{0j}-2h_{ij}Q^{j0}=0.$ We
have left the $Q$-term, because the statement, just as that about primary constraints
is valid both for affinely-invariant and Lorentz invariant models. Obviously, for
affine models the $Q$-term vanishes. Let us observe an interesting similarity to the
empty-space Einstein equations, where secondary constraints are related to
$R_{i}{}^{0}=0.$ Similarly, for the free electromagnetic field: $H^{0j}{}_{,j}={\rm
div} \overline{D}=0.$

As mentioned, discussion of the consistency of our model in terms of Dirac algorithm
would be extremely difficult. Nevertheless, one can show that our field equations are
not self-contradictory (this might easily happen in models invariant under
infinite-dimensional groups with elements labelled by arbitrary functions). Namely,
one can explicitly construct some particular solutions of very interesting geometric
structure. Of course, there is still an unsolved problem "how large" is the general
solution.

Analysing the structure of equations (\ref{11}) one can easily prove the following
\begin{theorem}
If field of frames $e$ has the property that its "legs" $e_{A}$
span a semi-simple Lie algebra in the Lie-bracket sense,
$[e_{A},e_{B}]=\gamma^{C}{}_{AB}e_{C},\ \gamma^{C}{}_{AB}=const,$
det$[\gamma^{C}{}_{DA}\gamma^{D}{}_{CB}]\neq 0,$ then $e$ is a
solution of (\ref{11}) for any GL$(n,{\bf R})$-invariant model of
$L$, in particular, for (\ref{12}).
\end{theorem}

Roughly speaking, this means that semisimple Lie groups, or rather
their group spaces are solutions of variational GL$(n,{\bf
R})$-invariant filed equations for linear frames. They are
homogeneous, physically non-excited vacuums of the corresponding
model. Fixing some point $a \in M$ we turn $M$ into semisimple Lie
group. Its neutral element is just $a$ itself, $e_{A}$ generate
left regular translations and are right-invariant. This gives rise
also to left-invariant vector fields ${}^{a}e_{A}$ generating
right regular translations,
$\left[e_{A},e_{B}\right]=\gamma^{C}{}_{AB}e_{C}, \
\left[{}^{a}e_{A},{}^{a}e_{B}\right]=-\gamma^{C}{}_{AB}{}^{a}e_{C},
\ \left[e_{A},{}^{a}e_{B}\right]=0.$ The tensor $\gamma_{ij}$
becomes then the usual Killing metric on Lie group; it is parallel
with respect to the teleparallelism connection $\Gamma_{\rm
tel}[e]$, $\nabla \gamma_{ij}=0.$ This means that
$\left(M,\gamma,\Gamma_{\rm tel}\right)$ is a Riemann-Cartan
space. For the general $e$ it is not the case. For semisimple
Lie-algebraic solutions there exists such a bilinear form $\eta$
on ${\bf R}^{n}$ (on $V$) that: $\gamma[e]=h[e,\eta]$ and
obviously $\eta_{AB}=\gamma^{C}{}_{DA}\gamma^{D}{}_{CB}.$ The
metric field $\gamma [e]$ has $2n$ Killing vectors
$e_{A},{}^{a}e_{A}$. Obviously, within the GL$(n,{\bf
R})$-invariant framework $\gamma[e]$ is a more natural candidate
for the space-time metric than $h[e,\eta]$. In the special case of
Lie-algebraic frames they in a sense coincide, but neither $\eta$
itself nor even its signature are a priori fixed. Instead, they
are some features, a kind of integration constants of some
particular solutions.

Let us mention, there is an idea according to which all fundamental physical fields
should be described by differential forms (these objects may be invariantly
differentiated in any amorphous manifold \cite{8,21,22,23}). Of particular interest
are the special solutions, constant in the sense that their differentials are
expressed by constant-coefficients combinations of exterior products of primary
fields.

The question arises as to a possible link between the above GL$(n,{\bf
R})$-frame\-work and the ideas of general relativity. For Lie-algebraic fields of
frames some kind of relationship does exist. Namely, if $\gamma_{ij}$ is the Killing
metric on a semisimple $n$-dimensional Lie group, $R_{ij}$ is its Ricci tensor and
$R$ --- the curvature scalar, then, as one can show \cite{12}
\begin{equation}
R_{ij}-\frac{1}{2}R\gamma_{ij}=-\frac{1}{8}(n-2)\gamma_{ij}.
\end{equation}
Rescaling the definition of the metric tensor on $M,\
g_{ij}=a\gamma_{ij},\ a={\rm const},$ we obtain
$R_{ij}-(1/2)Rg_{ij}=\Lambda g_{ij},\ \Lambda=-(n-2)/8a,$ and
these are just Einstein equations with a kind of cosmological
term. Therefore, at least in a neighbourhood of group-like vacuums
the both models seem to be somehow interrelated.

There is however one disappointing feature of affinely-invariant $n$-leg models and
their interesting and surprising Lie group solutions. Everything is beautiful for the
abstract, non-specified $n$. But for our space-time $n=4$ and there exist no
semisimple Lie algebras in this dimension. There are, fortunately, a few supporting
arguments:
\begin{enumerate}
\item We can try to save everything on the level of Kaluza-Klein universes of
dimension $n>4$. It is interesting that the $n$-leg field offers the possibility of
deriving the very fibration of such a universe over the usual four-dimensional
space-time as something dynamical, not absolute as in Kaluza-Klein theory. The
fibration and the structural group would then appear as features of some particular
solutions.
\item One can show that Lie-algebraic solutions exist in some sense for systems
consisting of the $n$-leg and of some matter field, e.g. the
complex scalar field $\Psi$. Lagrange tensor is then given by
$L_{ij}=(1-a \overline{\Psi}
\Psi)\gamma_{ij}+b\overline{\Psi}_{,i} \Psi_{,j}.$ Even if $e_{A}$
span a nonsimple Lie algebra, there are $(e,\Psi)$-solutions with
det$[L_{ij}]\neq 0$ and with the oscillating complex unimodulary
factor at $\Psi$ \cite{10,11}. The same may be done for
higher-dimensional multiplets of matter fields,
$L_{ij}=(1-a_{\overline{k} l}
\overline{\Psi}{}^{\overline{k}}\Psi^{l})\gamma_{ij}+
b_{\overline{k}l}\overline{\Psi}{}^{\overline{k}}{}_{,i}\Psi^{l}{}_{,j}.$
\item In dimensions "semisimple plus one" (e.g. $4$) there exist
also some geometric solutions with the group-theoretical
background. They are deformed trivial central extensions of
semisimple Lie groups \cite{21}.
\end{enumerate}

Let us describe roughly the last point. We fix some Lie-algebraic
$n$-leg field $E=(\ldots,E_{A},\ldots)=(E_{0},\ldots,E_{\Sigma},
\ldots)$, where $A=\overline{0,n-1}$, $\Sigma=\overline{1,n-1}$,
and the basic Lie brackets are as follows:
$\left[E_{0},E_{\Sigma}\right]=0,\
\left[E_{\Sigma},E_{\Lambda}\right]=E_{\Delta}C^{\Delta}{}_{\Sigma
\Lambda}, $ and ${\rm det}[C_{\Lambda \Gamma}]:={\rm
det}\left[C^{\Sigma}{}_{\Lambda \Delta} C^{\Delta}{}_{\Gamma
\Sigma}\right]\neq 0.$ In adapted coordinates
$(\tau,x^{\mu})=(x^{0},x^{\mu})$ (where $\mu=\overline{1,n-1}$) we
have $E_{0}=\partial/\partial \tau,\
E_{\Sigma}=E^{\mu}{}_{\Sigma}(x)\partial/\partial x^{\mu}.$ The
dual co-frame $E=(\ldots,E^{A},\ldots)=(E^{0},\ldots,E^{\Sigma},
\ldots)$ is locally represented as: $E^{0}=d \tau,\
E^{\Sigma}=E^{\Sigma}{}_{\mu}(x)dx^{\mu},\
E^{\Sigma}{}_{\mu}E^{\mu}{}_{\Lambda}=\delta^{\Sigma}{}_{\Lambda}.$
The corresponding Lie algebra obviously is not semisimple. But we
can construct new fields of frames $e$ or $'e$ given respectively
by $e=\rho E,\ 'e_{0}=E_{0},\ 'e_{\Sigma}=\rho
E_{\Sigma}=e_{\Sigma},$ where $\rho$ is a scalar function such
that $e_{\Sigma}\rho =E_{\Sigma}\rho=0$, i.e. in adapted
coordinates it depends only on $\tau,\ \partial \rho/\partial
x^{\mu}=0$.
\begin{theorem}
For any $\rho$ without critical points, both $e$ and $\ 'e$ are
solutions of any GL$(n,{\bf R})$-invariant and generally covariant
equations (\ref{11}). In both cases $\gamma[e]=\gamma['e]$ is
stationary and static in spite of the expanding (contracting)
behaviour of $e,\ 'e$.
\end{theorem}

If the Lie algebra spanned by $(E_{1},\ldots,E_{n-1})$ is of the
compact type, then $\gamma[e]$ is normal-hyperbolic and has the
signature $(+-\ldots -)$ with respect to the nonholonomic basis
$(E_{0},\ldots,E_{\Sigma},\ldots)$, thus the $\tau$-variable and
coordinates $x^{i}$ have respectively time-like and space-like
character. The above function $\rho$ is a purely gauge variable
and in appropriately adapted coordinates:
\begin{equation}\label{13}
\gamma[e]=\gamma['e]=dx^{0}\otimes dx^{0}+{}_{(n-1)}\gamma_{\alpha
\beta}(x^{\kappa})dx^{\alpha}\otimes dx^{\beta},
\end{equation}
where $x^{0}:=\pm \sqrt{(n-1)}{\rm ln}(x^{0}/\delta)$, $\delta$ is
constant, ${}_{(n-1)}\gamma_{\alpha \beta}=4S^{\kappa}{}_{\lambda
\alpha}S^{\lambda}{}_{\kappa \beta}$. Obviously, in all formulas
the capital and small Greek indices, both free and summed run over
the "spatial" range $\overline{1,n-1}$ (conversely as in the
usually used notation).

Another, coordinate-free expression: $\gamma[e]=(n-1)\left(d\rho /d\tau
\right)^{2}e^{0}\otimes e^{0}+\rho^{2}C_{\Lambda \Sigma}\\$$e^{\Lambda}\otimes
e^{\Sigma}.$ With such solutions $M$ becomes locally ${\bf R}_{\rm time}\times G_{\rm
space}$, $G$ denoting the $(n-1)$-dimensional Lie group with structure constants
$C^{\Delta}{}_{\Lambda \Sigma}$. The above metric $\gamma$ has $(2n-1)$ Killing
vectors; one time-like and $2(n-1)$ space-like ones, when $G$ is compact-type. This
is explicitly seen from the formula (\ref{13}), or its coordinate-free form
$\gamma=(n-1)\left(d{\rm ln} \rho /d \tau \right)^{2}E^{0}\otimes E^{0}+C_{\Lambda
\Sigma}E^{\Lambda}\otimes E^{\Sigma}.$ If we introduce spinor fields, then in their
matter Lagrangians we must use the Dirac-Einstein metric $h[e,\eta]$ with $\eta$ of
the form: $\eta_{00}=\beta={\rm const},\ \eta_{0\Lambda}=0,\ \eta_{\Lambda
\Sigma}=C_{\Lambda \Sigma}$. This metric is subject to the cosmological expansion
(contraction) known from general relativity, e.g. $h['e,\eta]$ in its spatial part
expands according to the de Sitter rule. Therefore, in spite of stationary-static
character of $\gamma$, the test spinor matter will witness about cosmological
expansion (contraction). This may be an alternative explanation of this phenomenon.
If $n=4$ there are the following Lie-algebraic-expanding vacuum solutions: ${\bf
R}\times {\rm SU}(2)$ or ${\bf R}\times {\rm SO}(3,{\bf R})$ with the
normal-hyperbolic signature $(+\ -\ -\ -)$, the plus sign related to $E_{0}$. There
are also solutions of the form ${\bf R}\times {\rm SL}(2,{\bf R})$, ${\bf R}\times
\overline{{\rm SL}(2,{\bf R})}$; they have the signature $(+\ +\ +\ -)$; now the
time-like contribution has to do with the "compact dimension" of SL$(2,{\bf R})$,
whereas the mentioned "expansion" holds in one of spatial directions. It is seen that
our GL$(n,{\bf R})$-models in a sense distinguish both the normal-hyperbolic signature
and the dimension $n=4$, just on the basis of solutions of local differential
equations. In any case, something like the $\eta$-signature of standard tetrad
description is not here introduced by hands.

Finally, let us observe that one can speculate also about another cosmological
aspects of our model. In generally-relativistic spinor theory one uses the Dirac
amplitude, tetrad and spinor connection (or affine Einstein-Cartan connection) as
basic dynamical variables. The corresponding matter (Dirac) Lagrangian is locally
SO$(1,3)$- or rather SL$(2,{\bf C})$-invariant. The same concerns gravitational
Lagrangian for the tetrad and spinor connection either in Einstein or in
gauge-Poincar\'{e} form. The idea was formulated some time ago that the true
gravitational Lagrangian should contain a term which is only globally invariant under
internal symmetries. Additional tetrad degrees of freedom were then expected to have
something to do with the dark matter, at least in a part of it \cite{3,4}. Our
GL$(4,{\bf R})$-models would be from this point of view optimal.

Finally, let us notice that our "expanding" Lie solutions for
dimensions "semisimple plus one" might be cosmologically
interpreted as the motion of cosmical relativistic fluid
($e_{0}$-legs of the tetrad) with internal affine degrees of
freedom ($e_{\Sigma}$-legs). This would be something like the
relativistic micromorphic continuum \cite{20,21}.
\bigskip

\small
\noindent {\bf Acknowledgements} The author is greatly indebted to
Colleagues from Cinvestav, first of all to Professor Maciej
Przanowski, for invitation to the conference and for their cordial
hospitality in Mexico City. The research itself and participation
were also partly supported by the Polish Committee of Scientific
Research (KBN) within the framework of the grant 8T07A04720.
\tiny

\end{document}